\documentclass[a4paper]{jpconf}

\begin{document}
\catcode`\@=11
\immediate\write\@auxout{\string\bibstyle{iopart-num}}
\catcode`\@=12
\title{Source of the observed thermodynamic arrow%
\footnote{For the proceedings of the conference DICE2008, Castiglioncello, Italy.}}
\author{L. S. Schulman}
\address{Physics Department, Clarkson University, Potsdam, New York 13699-5820, USA}
\ead{schulman@clarkson.edu}

\begin{abstract}  
The puzzle of the thermodynamic arrow of time reduces to the question of how the universe could have had lower entropy in the past. I show that no special entropy lowering mechanism (or fluctuation) is necessary. As a consequence of expansion, at a particular epoch in the history of the universe a state that was near maximum entropy under the dominant short range forces becomes extremely unlikely, due to a switchover to newly dominant long range forces. This happened at about the time of decoupling, prior to which I make no statement about arrows. The role of cosmology in thermodynamics was first suggested by T. Gold.
\end{abstract}

The thermodynamic arrow of time, basically the second law of thermodynamics, is often said to be one of the mysteries of nature, standing in stark contrast to the near time-symmetry of other dynamical laws accepted in physics.

This article is less ambitious than other recent discussions of the problem, for example, Ref.~\cite{wald}. I only want to explain the observed phenomenon in terms of other known and accepted pieces of physics. I will not go to the very early universe (not even the first millennium), nor will I invoke unobserved phenomena. What I do assert is that there is less mystery than is alleged. The essential features of this argument have already appeared both in my own writings \cite{timebook, compromisedarrowvaillant, timelecturesarrows} and in the writings of others \cite{penrose}.\footnote{See p.\ 135 in \cite{timebook}, p.\ 358 in \cite{compromisedarrowvaillant}, Sec.\ 7 in \cite{timelecturesarrows} and p.\ 705 in \cite{penrose}.} In the present article I will review and extend these arguments. 

As foreshadowed, the explanation will involve cosmology. Notwithstanding the asserted limited scope, the claim is far from trivial: it relates local thermodynamics to the universe at large, a connection that is neither obvious nor intuitive. This leads me to a secondary objective of this presentation, namely to trace the origin of these ideas to their proper source, namely Thomas Gold, who in at least some of the current literature is either ignored or misquoted
.\footnote{I do not allude to Ref.\ \cite{wald}. The \textit{mis}quotation is the claim that Gold talked about a universe with expansion and contraction, with reversal of the arrow at the midpoint. He did talk about a contracting universe and the idea that the arrow would be reversed in it. But both at once---not. Now I happen to like that idea, but whatever he thought of it, he was adamant that it should not be attributed to him (personal communication). I don't know the first of these careless attributions, but at least some of the misquotation comes from those who used earlier misquotations rather than going to the easily available primary source~\cite{gold1}.}

Around 1960, 30 years after the discovery of the expansion of the universe, Gold suggested \cite{gold1} that the thermodynamic arrow was a consequence of that expansion. That this was a significant step can be seen in an earlier article by Wheeler and Feynman in which they start from a time-symmetric electrodynamics and recover the usual time-asymmetric theory by means of their ``absorber theory'' \cite{wheelerfeynman1A}. Their work deals in an essential way with the issue of irreversibility (for example in presenting a framework for Dirac's calculation of radiation reaction \cite{dirac}) and in some of their attempts to establish the absorber theory Wheeler and Feynman make reference to the universe at large. Nevertheless, the essential feature, that the universe is \textit{expanding}, comes nowhere in their article. Ref.\ \cite{wheelerfeynman1A} even mentions discussion with Einstein. This was in the 1940's and expansion did not seem the route to solving this great puzzle. For context it is interesting to read the record of a conference convened by Gold and Bondi \cite{goldbook} in 1963. Some participants were sympathetic to the idea that expansion could play a role (e.g., \cite{hoyle}); many were not. 

In any case, the arguments of Gold indicated ways in which large scale expansion could be felt locally. He reasoned \cite{gold1} that any local arrow is imposed by an outside influence, with the ultimate outside influence being the expansion of the universe, manifested through outgoing wave electromagnetic boundary conditions. Although I found his thesis attractive, I felt \cite{correlating} that a proper presentation of his argument required time-symmetric boundary conditions. In the present article a slightly different approach is taken, based on actual observation of earlier states of the universe, combined with known dynamical processes.

It was realized long ago by Boltzmann that the problem of justifying the arrow was not so much showing that the entropy of isolated systems increased (accomplished with controversy by his H-theorem), but explaining why there was low entropy in the past. His justification \cite{boltzmann} was that there was an enormous ``fluctuation.'' For most of the universe and for most of its history it is in equilibrium, but at some point in our past there was a fluctuation leading to our present situation. It is hard to argue with this assertion; it is also hard to believe. See \cite{timebook}, notes to Sec.\ 4.0, and many other discussions.

I now argue that what we already know about the history of the universe provides that low entropy, and it occurs by virtue of the expansion
.\footnote{A note of caution must be sounded, even when arguing from an early low entropy state. Once a particular time has been selected, symmetric dynamics \textit{away from} that time, in either direction, will tend to increase (or not decrease) entropy. This is the first conundrum of the Boltzmann H-theorem. As pointed out in \cite{correlating, timebook}, the use of \textit{initial} conditions generally provides an arrow of time with little else required. Thus to make a non-circular argument, one can take these low entropy conditions as \textit{two-time} boundary conditions. Then if entropy increases towards the middle the desired result is established. Such arguments appear in~\cite{timebook}. For the reasoning in the present article this is not a problem, since there is a substantial change in dominant dynamics at the time of decoupling. Moreover, I make no claim about having or not having an arrow at earlier epochs.}

Consider the era of recombination or decoupling. For that epoch, we have an excellent image of the universe, namely that provided by the cosmic background radiation. It shows, for the luminous matter, a tremendous level (1 part in 10$^5$) of uniformity. This is reasonable. Under short range forces, for example the usual intermolecular interactions, the most likely state, that of maximum entropy, is uniformity. Forces were short range prior to recombination because the system was basically a plasma. Although electromagnetic forces are long range, because of the presence of uncancelled positive and negative charges there was screening of the electromagnetic forces, making them effectively short range. Of course gravitational forces were also present, but for luminous matter they were negligible. For dark matter, it is generally felt that at that time there must already have been some level of non-uniformity, but that does not affect our argument. Similarly, there is no evidence for the presence of black holes. To summarize the significant observation: the luminous matter was in equilibrium under its dominant forces and that equilibrium was a uniform state. 

With the advent of recombination, charged particles combined to form neutral objects, H atoms, and the universe became transparent. Atom-atom forces were also negligible since particle separations were on the order of mm. Gravity now became the dominant force. Under gravity, uniformity is \textit{not} the most likely state.\footnote{As an alternative way of characterizing a state as ``unlikely'' I'll say that its entropy is low. Nevertheless, entropy is problematic when long range forces are involved. Binney and Tremaine (\cite{binney}, p.\ 268)  give a Newtonian-gravity argument showing that in a galaxy entropy can increase without limit through rearrangements of the stars. I would also say that there is a more fundamental issue here. In the usual gas-with-short-range forces, one \textit{loses} information as the gas becomes homogeneous. Ergodicity is equivalent to a system's losing macroscopic observables, until the only one left is energy (or in some circumstances, momentum and angular momentum). With gravity, what may have been a homogeneous system begins to \textit{acquire} features; portions of the ``gas'' separate and new observables are born. However, for the purposes of discussion in the present article, I will use the word ``entropy'' as a shorthand for the likelihood or unlikelihood of states. But the kind of calculation in which one assigns a particular numerical value of entropy to a certain state (as one does in conventional thermodynamics) is not at all contemplated. Where there is common ground is in the consequences of a system being in a likely or an unlikely state. As for the usual low entropy situation, a system that can be identified as of low likelihood will progress to states of higher likelihood. In our case, the low likelihood state---under gravity---is uniformity.} On the contrary, matter tends to clump, in fact it seems to do so hierarchically. The uniform distribution that in an earlier era was the most likely configuration, now becomes unstable.

This then is the punch line: with no mechanism other than expansion, entropy has been lowered. Interestingly, it's not the state that changes, but the dominant forces. What was maximum entropy in one regime becomes low entropy in another. And this happens because of expansion, since it's expansion that induces the cooling, causing in turn the formation of H atoms, and finally leading to a world in which the newly dominant force---gravity---controls a state that is unlikely from a gravitational perspective. This then is Gold's thesis, although the detailed arguments are different. 

Getting from this state of disequilibrium to our experienced arrow of time is qualitatively straightforward, although quantitatively challenging. Under gravitational forces there is a continuing emergence of ever greater levels of non-uniformity. Eventually this gives rise to star formation and the release of nuclear energy. This in turn leads to the negentropy flow that permits the preparation of local low entropy states. In our case, this negentropy flow takes the form of 6000$\,$K photons that allow us to remain out of equilibrium on earth.

Establishing this picture quantitatively is the objective of the fields of structure formation and stellar evolution. Puzzles in the former area suggest that, as remarked above, a degree of structure was already present before recombination. Presumably, this would ultimately have led to the high degree of non-uniformity that we see today, but it would not have done so on the observed time scale (about 13 billion years). This presumption, however, applies to a fictional world sans expansion: in our world, expansion implies decoupling, the trigger for the relatively rapid development of non-uniformity.

The foregoing argument says nothing about a thermodynamic arrow \textit{before} recombination. Perhaps there was none or perhaps it was there because of even earlier physical processes.\footnote{A fascinating essay by Misner \cite{misner} suggests that time should be measured logarithmically in the early universe (the way one often thinks about low temperatures), so that there could have been a succession of eras with arrows and perhaps with heat deaths and revivals under mechanisms such as described here.} We offer no opinion on this issue, since this paper is devoted to explaining the thermodynamic arrow we now experience.\footnote{\label{footnote:arrowrate}Having or not having an arrow need not be a yes-or-no issue. For example, one would say that a gas \textit{at} equilibrium has no arrow, while a gas released from a small volume surely does. This suggests that rate of change of entropy be a quantitative measure of an arrow. However, this too has pitfalls. For a metastable system described by stochastic dynamics the derivative may be small (fixed by the second largest eigenvalue of the matrix of transition probabilities \cite{firstorder}) while for any particular exemplar the apparent changes may be great (after a nucleation event) or seemingly nonexistent (in the absence of such an event). For the systems discussed here there are two differences from the scenario just mentioned. First, with long range forces entropy is problematic. Second, the nucleation events are \textit{not} sudden and local, but may occur at large distance scales. As such, on smaller scales (where entropy may be defined in a local sense) there will be no perceivable arrow.}

Finally a comment about the term ``adiabatic'' when used in connection with the post-decoupling expansion of the universe. I mention this because in the discussion in \cite{timebook} the expansion is characterized as rapid. Certainly, adiabatic, suggesting a slow process, is appropriate when considering the ability of the photon gas to keep up with the expansion. However, the clumping process, that which ultimately leads to stars, etc., is far slower than the expansion. In other words, as the universe expands, it does not manage to reach the state of greatest likelihood associated with gravity-dominated dynamics (which may not exist).

In this article I have given qualitative arguments. To what extent can greater precision be obtained? This could be in the form of better estimates of inhomogeneity, or perhaps model systems exhibiting the features I postulate. Frankly, for my purposes I don't think this is necessary, and for two reasons. First, behind my assertions are decades of experience and calculations by many, many people. Gravity makes things clump; of course much more than that is known, time scales under various conditions and the like, but this is greater precision than I need. Second, I am not predicting inhomogeneity; I am observing it. The evident property of gravity to head toward such a configuration is essential to my argument, but for that argument it is enough simply to observe that uniformity and gravity are at odds.

Similar remarks apply to models, but discussing the phenomenon in phase transition terms is instructive. Consider a Hamiltonian with both short and long range interparticle forces. If there is short-range repulsion and it is bounded in magnitude, there may very well may be no stable state of the system. However, if the long-range force is much weaker than the other, there can be a long-lived (uniform density) metastable phase. The passage out of this metastable state can be considered a phase transition,\footnote{This phase transition would not be well described using the conventional notion of a breakdown in analyticity. However, except for mean field models (e.g., the van der Waals gas), analyticity methods fail to describe metastable states, and more general concepts, relying on time-scale differences, are needed~\cite{firstorder}.} from something that looks like equilibrium under short-range forces to something for which the dynamics never comes to equilibrium. Note that I am not saying that the actual universe suddenly passed through such a transition. What I mean by ``phase transition'' is the dramatic dependence of system behavior on a parameter. In our case we could take the parameter to be the strength of the short range force. The lifetime of the metastable phase would then drop rapidly as the short range force weakened. Thus, if we considered the plasma (sans dark matter) that preceded recombination but in a \textit{non\/}expanding universe, the lifetime for gravity to take over and cause clumping would be extremely long.\footnote{Moreover, without expansion the dark matter inhomogeneities that developed beginning with matter dominance would require a much extended time scale before they affected the luminous matter, assuming it remained a plasma.} However, if the short range coupling constant is reduced (as it is, because of cooling and expansion), then that lifetime has the dramatic reduction that one associates with a phase transition.

I review what has or has not been explained here. I have \textit{not} explained why the universe was in a metastable state prior to recombination, a state in which not only the inhomogeneity of the gravitational dominance was absent, but the ultimate inhomogeneity of black holes was also absent. What I \textit{have} explained is how the state that is in fact observed leads to the second law of thermodynamics, as we now see it. My point is thus the connection or correlation of two observed phenomena. In one sentence I restate the explanation: Expansion-induced cooling shifted the strength-balance of long and short range forces, allowing nucleation of a matter distribution that is clumpy and appears to have no ultimate equilibrium. Or more loosely speaking, what \textit{was} high entropy with respect to the dominant forces becomes low entropy by virtue of an expansion-induced change in which forces dominate.

\def\texttilde{$\scriptstyle\sim$}

As declared at the outset, the scope of the explanation offered here is more modest than that of other recent attempts. There is good reason to be conservative. The field of cosmology undergoes major revisions of its paradigms with impressive frequency.\footnote{Examples come to mind: adjusting Hubble's distance scale by a factor two, dark matter, inflation, dark energy. Although at least one cosmologist (www.hep.phys.soton.ac.uk/\texttilde sfk/cosmology.ppt) seems to think that Landau's adage about cosmologists, ``often in error, never in doubt,'' has been abrogated, I don't \hbox{share that opinion.}} A great deal of modern cosmology is concerned with properties of black holes, and for the topic at hand focuses particularly on their entropic properties. As far as I know there is no experimental or observational evidence on this matter, nor is there a shred of evidence, beyond theoretical conviction, for the existence of radiation from black holes. 

To summarize, the existence of the observed thermodynamic arrow of time has been traced to observed and non-controversial features of the not-too-early universe. Given the state observed at that time, coupled with the expansion of the universe, one obtains the contemporary arrow of time.

\section*{Acknowledgments}
I am grateful to the organizers of the DICE2008 conference for the opportunity to present these ideas. At the conference itself I gave additional background with emphasis on the need to avoid ``initial conditions prejudice,'' in which the natural structure of language can lure one into circular arguments. This part of my lecture has been omitted from the present article and is mostly covered in \cite{timebook}. More recent work of mine on this is \cite{opposite} in which the compatibility of interacting opposite-arrow regions is shown, as well as \cite{causalityA} in which macroscopic causality loses its status as an independent property of nature and is shown to be a consequence of general entropy increase, i.e., the thermodynamic arrow.

The article itself was significantly enhanced by extended discussions with Amos Ori and Marco Roncadelli. I am also grateful to Bernard Gaveau and Huw Price for helpful discussions. Much of this interaction took place in the ``Advanced Study Group,'' \textit{Time: quantum and statistical mechanics aspects}, of the Max Planck Institute for the Physics of Complex Systems, in Dresden. This work was partly supported by the United States National Science Foundation grant PHY~05~55313. I am also grateful to the Lewiner Institute for Theoretical Physics for partial support during a visit to the Technion.
\def\texttilde{$\scriptstyle\sim$}
\section*{References}\vskip-40pt
\providecommand{\newblock}{}

\end{document}